\documentclass[12pt]{article}
\usepackage{url}
\usepackage{epsfig,epsf}
\usepackage{graphicx,graphics}
\usepackage{amssymb,amsmath}
\usepackage{verbatim,enumerate}
\usepackage{verbatim,multicol,color}
\usepackage{multirow}
\usepackage{booktabs}
\usepackage{multicol}
\usepackage[english]{babel}
\usepackage{blindtext}
\usepackage{subfig}
\usepackage{natbib}
\usepackage{float}

\def\log{\hbox{log}}

\def\boxit#1{\vbox{\hrule\hbox{\vrule\kern6pt
          \vbox{\kern6pt#1\kern6pt}\kern6pt\vrule}\hrule}}

\def\bse{\begin{eqnarray*}}
\def\ese{\end{eqnarray*}}
\def\be{\begin{eqnarray}}
\def\ee{\end{eqnarray}}
\def\bq{\begin{equation}}
\def\eq{\end{equation}}
\def\bse{\begin{eqnarray*}}
\def\ese{\end{eqnarray*}}

\newcommand{\bsy}{\boldsymbol}

\newcommand{\mbf}{\mathbf}

\begin{document}
\thispagestyle{empty}
\baselineskip=28pt
\vskip 5mm
\begin{center} {\Large{\bf A Multi-Resolution Spatio-Temporal Model for Brain Activation and Connectivity in fMRI Data}}
\end{center}

\baselineskip=12pt
\vskip 5mm

\begin{center}
\large
Stefano Castruccio{\footnotemark[1]}, Hernando Ombao{\footnotemark[2]} and Marc G. Genton{\footnotemark[3]}
\end{center}

\footnotetext[1]{
\baselineskip=10pt School of Mathematics \& Statistics, Newcastle University, Newcastle Upon Tyne, NE1 7RU United Kingdom. E-mail: stefano.castruccio@ncl.ac.uk}

\footnotetext[2]{
\baselineskip=10pt Department of Statistics and Department of Cognitive Sciences, University of California, Irvine, CA 92697, United States. E-mail: hombao@uci.edu}

\footnotetext[3]{
\baselineskip=10pt Computer, Electrical and Mathematical Sciences and Engineering Division, King Abdullah University of Science and Technology (KAUST), Thuwal 23955-6900, Saudi Arabia. E-mail: marc.genton@kaust.edu.sa}
\baselineskip=16pt
\vskip 4mm
\centerline{\today}
\vskip 6mm

\begin{abstract}
Functional Magnetic Resonance Imaging (fMRI) is a primary modality for studying brain activity. Modeling spatial dependence of imaging data at different scales is one of the main challenges of contemporary neuroimaging, and it could allow for accurate testing for significance in neural activity. The high dimensionality of this type of data (on the order of hundreds of thousands of voxels) poses serious modeling challenges and considerable computational constraints. For the sake of feasibility, standard models typically reduce dimensionality by modeling covariance among regions of interest (ROIs) -- coarser or larger spatial units -- rather than among voxels.
However, ignoring spatial dependence at different scales could drastically reduce our ability to detect activation patterns in the brain and hence produce misleading results. To overcome these problems, we introduce a multi-resolution spatio-temporal model and a computationally efficient methodology to estimate cognitive control related activation and whole-brain connectivity. The proposed model allows for testing voxel-specific activation while accounting for non-stationary local spatial dependence within anatomically defined ROIs, as well as regional dependence (between-ROIs). Furthermore, the model allows for detection of interpretable connectivity patterns among ROIs using the graphical Least Absolute Shrinkage Selection Operator (LASSO). The model is used in a motor-task fMRI study to investigate brain activation and connectivity patterns aimed at identifying associations between these patterns and regaining motor functionality following a stroke.
\end{abstract}

\baselineskip=16pt

\par\vfill\noindent

\par\medskip\noindent
{\bf Short title}: A Multi-Resolution Model for fMRI data

\clearpage\pagebreak\newpage \pagenumbering{arabic}
\baselineskip=26pt

\section{Introduction}

Detecting and understanding significant patterns of brain activity is among the most important challenges of contemporary science. To this end, functional Magnetic Resonance Imaging (fMRI) data has been at the center of neuroscience investigations for the last twenty years. fMRI measures brain activity by detecting changes in neural activity associated with blood flow using the contrast between deoxygenated hemoglobin (which is paramagnetic) to oxygenated hemoglobin (which is diamagnetic) in localized spatial volumes called \textit{voxels}. An fMRI scan produces a highly spatially resolved brain imaging data set. The statistical challenge is to develop a model that is able to detect voxels and regions of interest (ROIs) that are activated during a performed task, to understand how the ROIs function together by incorporating information from all locations (there may be as many as 150,000 voxels over the entire brain volume for each scan) and to describe the spatial dependence while still allowing for scalable inference. 
 
In this paper, we report the results of a collaboration with the stroke rehabilitation center at UC Irvine on a project that aims to identify brain activation and connectivity patterns during the execution of a motor task (e.g., hand grasping). Throughout this work, connectivity will be defined as the conditional dependence across ROIs.

The simplest approach in analyzing fMRI data is to fit a linear model (often termed \textit{general linear model} in the brain imaging literature) for voxel-specific (or ROI-specific) time series. However, this approach does not account for spatial dependence across voxels or between ROIs, and hence could potentially result in misleading conclusions. It is well known that if spatial dependence is ignored, then the uncertainty of the estimators is assessed incorrectly, thus inducing inflated Type I error rates \citep{dub88} when testing for significance. While it is possible to partially adjust the analysis for spatial correlation (e.g., by post-processing the data via spatial smoothing), our goal in this work is to develop a comprehensive model that defines the spatio-temporal dependence (both at local and global scales) of a complete fMRI data set which is necessary to fully and correctly account for these effects. Although most researchers acknowledge that taking spatial dependence into account is important, the key obstacle has been the seemingly insurmountable computational cost. Here, we develop a computationally efficient algorithm that overcomes this major limitation in spatio-temporal models for fMRI data.

The earliest approaches to modeling fMRI data focused only on within-voxel temporal correlation, either ignoring spatial correlation or imposing unrealistic constraints to reduce the computational burden associated with the high spatial dimensionality of fMRI data \citep{Worsley1995,Locasio1997,Bullmore2001}. Seminal works on statistical methods for fMRI data (the random field theory developed in \cite{Worsley1992} and \cite{Worsley1995}) indirectly accounts for spatial correlation by assuming that the voxel-specific test statistics (e.g., the $t$-statistic or $F$-statistic) are realizations of some random field, and then imposing a thresholding approach based on the number of resolution elements (\textit{resels}, \cite{Worsley1992}). The thresholding depends on the spatial smoothness, which is assumed to be isotropic across the brain. 

\cite{bow05} proposed a two-stage hierarchical Bayesian approach to first estimate activation patterns under the assumption of spatial independence and then to model the spatial dependence of the mean within regions. \cite{Bowman2007} and \cite{bow08} extended this work by allowing correlation for each voxel within a region, and \cite{der10} proposed a model to also account for temporal correlation between multiple experimental effects. The two-stage approach was the first rigorous framework to acknowledge spatial correlation and has given rise to a large body of literature on Bayesian models for fMRI data (see \cite{zha15} for a comprehensive review). Although this framework has been demonstrated to be flexible and to produce useful information for practitioners (e.g., posterior probability maps for activation), two main factors still present limitations to the development of spatio-temporal models for fMRI data. Firstly, dependence of activation patterns (or some differencing of them) assumes a Gaussian Markov random field in the mean structure, which is a natural choice given the gridded geometry of fMRI, and independent errors. This approach implies stationarity for voxels within the same ROI, an hypothesis which is in practice violated, as we show in this work. Secondly, inference is often limited to subsamples such as two-dimensional slices to reduce the dimensionality and consequently the computational time. Indeed, \cite{zha15} claim that `the large dimensionality of the data makes it impossible to model the entire 3D maps of the data at once.' Here, we demonstrate that it is possible to provide activation maps for the entire brain with locally anisotropic spatio-temporal models, provided that a suitable approximate inferential scheme is implemented.

An alternative approach to modeling spatial dependence was proposed by \cite{ka12}, where a spatio-spectral mixed-effects model that captures multi-scale spatial correlation (among ROIs and within ROIs) was proposed. By defining the model on the spectral domain (i.e., modeling the Fourier coefficients rather than the fMRI time series), their approach holds promise for scalability because the Fourier coefficients are approximately uncorrelated across different frequencies under a temporally stationary assumption. By incorporating voxel-specific and ROI-specific random effects, the model captures the spatial covariance structure both on a local level (where the local correlation
between voxels depends on their distance) and on a regional level (where the correlation
between regions is not forced to depend on distance) without reducing the analysis to 2D slices. However, the activation was assumed at the regional level, not at the voxel level, and the assumption of isotropy within a region is, as we show in this work, not appropriate. With a more sophisticated methodology to handle nonstationary spatial data and more powerful computing resources available, this assumption can and should be relaxed.

\cite{DegrasLindquist} developed a hierarchical model for voxel-specific and condition-specific activation and inference
in a multi-subject setting. The problem of simultaneously estimating the Hemodynamic Response Function (HRF) and
voxel activation was previously discussed in \cite{Makni2005} and \cite{Makni2008}.
The approach in \cite{DegrasLindquist} uses a set of B-spline basis functions to represent the HRF. The coefficients of these functions are allowed to vary between experimental conditions, across voxels in space and
over all subjects. Similarly, \cite{zha12,zha13,zha14} proposed different estimation strategies of the HRF, both in the parametric and semi-parametric setting for multiple subjects. The proposed models are flexible and the estimation-inference procedure is rigorously developed; however all of them still lack a definition of the spatial covariance which is able to capture the local nonstationary dependence. 

More recently, \cite{zhu14} and \cite{hyu14} proposed mixed effect models with spatially varying coefficients that allow for spatial discontinuities in Blood Oxygenation Level Dependent (BOLD) activation and spatial dependence. While these models allow for a rich and flexible structure in the mean function, they also require an explicit and interpretable definition of a functional structure allowing jumps, which is challenging to implement with a very large number of voxels. Also, the covariance structure was assumed to have either a low rank representation \citep{cre08} or to rely on Gaussian predictive processes \citep{ban08} which do no explicitly model nonstationarity and lead to loss of information when the spatial correlation is moderate or strong \citep{st14}. 

In this work, we propose a new model that overcomes the aforementioned limitations by introducing a spatially varying brain response, and an error structure which is locally nonstationary and also regionally descriptive of the connectivity across ROIs. Inference can be achieved within the context of our data example, comprising of more than 22 million observations, via a multi-resolution approach without resorting to subsampling either with 2D slices or by assuming a coarse-level activation structure. The isotropy assumption for dependence is also relaxed assuming instead a locally anisotropic model \citep{fu01} for each ROI. Subsequently, generalized shrinkage \citep{fie11} with the empirical covariance matrix is performed to allow sufficient flexibility in capturing high spatial frequencies. The regional dependence structure is estimated with a sparse inverse structure via graphical Least Absolute Shrinkage and Selection Operator (LASSO) \citep{fri08} which allows for visualization of connectivity patterns across ROIs, exploiting its interpretability as a graphical model \citep{mei06}. 

The Gaussian model that we propose allows for spatially varying coefficients and a multi-resolution (local and regional) error structure so that it is possible to realistically capture the complex spatio-temporal dependence across voxel-specific fMRI time series. Our model has two advantages. Firstly, a more realistic description of spatial dependence both for the mean and error structure allows for an improved inference when testing for activation, with less false positives than a model that assumes an overly simplistic dependence (or no dependence at all). This allows for a better detections of ROIs with a high degree of activity when the patient is performing the motor-task (although this approach can be generalized to other types of clinical trials). Secondly, a realistic model for regional dependence with a sparsity structure allows us to deduce an interpretable functional connectivity graph that provides information on how are the different ROI connected during the task. 

The remainder of the paper is organized as follows. Section \ref{s2ds} describes the data set, and Section \ref{mod_def} introduces the temporal, local and regional spatial structure of the spatio-temporal statistical model. Section \ref{mod_comp} provides a multi-resolution  inference scheme and discusses how distributed computing is instrumental in fitting the model to a data set of such scale. Section~\ref{sim_stud} presents two simulation studies to highlight the need for nonstationary models on the local scale, and Section \ref{s5disc} shows how a spatio-temporal model is able to detect more active voxels than a model with spatial independence and how these voxels correspond to interpretable patterns related to the task performed in the clinical trial. Section \ref{s6conc} concludes with a discussion and future directions of investigation.

\section{Experimental Design and fMRI Data}\label{s2ds}

\subsection{Experiment and goals}
Our motivating example comes from a clinical study from the neuro-rehabilitation laboratory of Dr. Steven C. Cramer, neurologist at the University of California at Irvine. The primary goal of the study was to investigate associations between motor functional deficits in stroke patients and brain
activation and connectivity. The group consists of all male, right-handed subjects between 18-35 years old. Our goal here is to develop a new model that explicitly takes into account the spatial dependence with a corresponding computationally efficient estimation algorithm and to demonstrate its feasibility for analyzing a single-subject full-brain voxel-level fMRI data set. We present here an analysis of a stroke affected individual. Although an extension to multi-subjects is possible, a rigorous approach must take into account a number of important issues, such as variation between subjects, which are beyond the scope of this work.

\begin{figure}[h]
  \centering
\includegraphics[scale=.3]{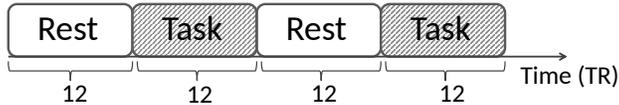}
   \caption[1]{Block design of the experiment for each of the three sessions performed by each subject of the clinical trial. The session comprises 48 time repetitions, each one consisting of a 2 second fMRI scan. `Task' corresponds to a hand grasp-and-release activity. }
  \label{real-design}
\end{figure}

In this experiment, there is a task and rest condition. During the task condition, the subjects perform a hand grasp-and-release movement task. The experiment was divided into three sessions, and each session had 48 consecutive scans, alternating between task and rest conditions three times, but always starting with the rest condition (see Figure \ref{real-design}). Therefore the total number of time points is $T = 144$.

\subsection{fMRI data and preprocessing}

The data were collected using a Philips Achieva 3.0T MRI whole-body scanner. The fMRI images were acquired using a T2*-weighted gradient-echo-planar imaging sequence with repetition time (TR) = 2000 ms, echo time (TE) = 30 ms, flip angle = 70, Field Of View (FOV) = $240\times240\times154$, slices = 31, and voxel size = $2\times2\times2$ mm$^3$.

Functional data from all sessions were preprocessed using SPM8 software (Wellcome Department of Imaging Neuroscience, www.fil.ion.ucl.ac.uk/spm). Preprocessing steps included realignment to the first image, coregistration to the mean image and normalization to the standard template.

To obtain anatomically defined ROIs, we used the Anatomical Automatic Labeling (AAL) atlas, which gives anatomical parcellation of the whole brain into 45 regions in each hemisphere \citep{tzo02}. These anatomical regions are
listed in Table S1 in the supplement. The number of voxels included in the AAL are approximately 150,000 for each scan, for a total of approximately 22 million data points. Voxel-specific fMRI time series were then extracted based on the subject-specific AAL-derived brain parcellation.


\section{The Statistical Model for fMRI data}\label{mod_def}

Throughout this work, $Y_{v}(t)$ denotes the fMRI time series at voxel $v$ and $r_v$ denotes the ROI of voxel $v$.
At any given time, we assume that there are two types of stimuli: active and rest. Let $S_1(t)$ be the indicator function during `task' in the block design in Figure \ref{real-design} and $S_2(t)=1-S_1(t)$ the indicator for `rest', so that exactly one condition is present for each time $t$.

We further assume that the HRF, denoted $h(t)$, is known and is common across all voxels. Voxel-specific and ROI-specific HRFs have been developed (see, e.g., \cite{DegrasLindquist}) by assuming that the HRF is the difference between two gamma functions or has a Poisson distribution \citep{zha15}, and allowing spatially varying parameters. We use the canonical HRF from the Statistical Parametric Mapping software to produce a BOLD response associated with each of the two stimuli. The BOLD response for the active condition, denoted $X_1(t)$, is the convolution of the HRF, $h(t)$, with the active stimulus indicator, $S_1(t)$: $X_1(t) = (h \ast S_1)(t)$. Similarly, the BOLD response for the rest condition is $X_2(t) = (h \ast S_2)(t)$.

\subsection{Voxel-wise activation}\label{1voxel}

Let $\mbf{Y}(t)=\{Y_{1}(t),\ldots,Y_{V}(t)\}^{\top}$ be the fMRI intensity at time $t$ for all voxels $v=1,\ldots, V$. We assume the following standard model for fMRI:
\begin{equation}\label{voxel_act}
\mbf{Y}(t) = \bsy{\beta}_{0}+\sum_{j=1}^J \bsy{\beta}_{j} I(t \in C_{j})+\bsy{\beta}_{J+1} (t \text{ mod } T/3)+\bsy{\beta}_{J+2} X_1(t) + \bsy{\beta}_{J+3} X_2(t) + \bsy{\varepsilon}(t),
\end{equation}
where $C_j$ indicates the $j$th session; $\bsy{\beta}_{j}=(\bsy{\beta}_{j;1},\ldots,\bsy{\beta}_{j;V})^{\top}$ for $j=0,\ldots,J+3$ are the covariates that are allowed to change at each voxel. Specifically $\bsy{\beta}_{0}$ is the intercept, $\bsy{\beta}_{1},\ldots,\bsy{\beta}_{J}$ allow for a changing mean for the sessions (in our case $J=2$, the third session mean is equal to zero for identifiability) while $\bsy{\beta}_{J+1}$ accounts for a temporal effect. Here, $\bsy{\beta}_{J+2}$ represents the linear contribution of the BOLD response, $X_1(t)$, while $\bsy{\beta}_{J+3}$ accounts for $X_2(t)$. The proposed model could also be extended to $M$ stimuli, with terms $\bsy{\beta}_{J+m+1} X_m(t)$ for $m=1,\ldots,M$ and $X_m(t)=(h \ast S_m)(t)$ with $S_m(t)$ being the indicator for the $m$th stimulus.

The noise $\bsy{\varepsilon}(t)$ is modeled as a vector autoregressive process of order 2 (VAR(2)). Note that the most common analyses
use the VAR of order $1$ or $2$, and even such low VAR orders already have $2V$ or $3V$ unknown parameters, respectively. Higher orders may be possible with a small number of ROIs but could be potentially computationally
challenging at the voxel level \citep{Bowman2007,DegrasLindquist}. Thus, $\bsy{\varepsilon}(t)$ is written as
\begin{equation}
\bsy{\varepsilon}(t) = \bsy{\Phi}_1 \bsy{\varepsilon}(t-1)+\bsy{\Phi}_2 \bsy{\varepsilon}(t-2)+\mbf{S}\{\bsy{\Omega}\mbf{H}_1(t)+(\mbf{I}_V-\bsy{\Omega})\mbf{H}_2(t)\},
\label{AR2}
\end{equation}
where $\mbf{I}_V$ is the identity matrix of size $V\times V$ and $\bsy{\Phi}_1=\{\phi_{1;v}\}, \bsy{\Phi}_2=\{\phi_{2;v}\}$ are $V\times V$ diagonal matrices with coefficients representing the autoregressive components of $\bsy{\varepsilon}(t)$. Here $\mbf{S}=\{\sigma_v\}$ is a diagonal matrix with voxel-wise standard deviations, and $\bsy{\Omega}\mbf{H}_1(t)+(\mbf{I}_V-\bsy{\Omega})\mbf{H}_2(t)$ is the vector of unscaled innovations. We assume that $\mbf{H}_u(t)\sim \mathcal{N}(\mbf{0},\bsy{\Sigma}_u)$ for $u=1,2$ where $\bsy{\Sigma}_u$ are correlation matrices and that $\mbf{H}_1(t)$ is independent from $\mbf{H}_2(t)$. The vector $\mbf{H}_1(t)$ controls the local (voxel-specific) scale dependence: its covariance $\bsy{\Sigma}_1$ is a block diagonal matrix, where each block $\bsy{\Sigma}_{1,r}$ corresponds to the dependence within ROI $r$. The vector $\mbf{H}_2(t)$ controls the regional scale dependence, representing the ROI specific effect with correlation $\bsy{\Sigma}_2$. $\bsy{\Omega}$ is a $V\times V$ diagonal matrix with diagonal elements $\omega_r \in [0,1]$ for each ROI $r$, which represents the relative contribution of the local random effect $\mbf{H}_1(t)$ compared to the regional random effect $\mbf{H}_2(t)$. We denote by $\mbf{H}_{1}(v)$ and by $\mbf{H}_{2}(r_v)$ the value of $\mbf{H}_{1}$ and $\mbf{H}_{2}$ at voxel $v$, respectively (the time index is removed for simplicity).

\subsection{Modeling intra-ROI dependence}\label{2roi}

We assume that cov$\{\mbf{H}_{1}(v),\mbf{H}_{1}(v')\}=0$ if $r_v\neq r_{v'}$; that is, voxels in different ROIs have no dependence through $\mbf{H}_1$. If the voxels belong to the same ROI, that is, if $r_{v}=r_{v'}=r$, a model for the spatial dependence must be defined. Previous works \citep{ka12,ka13} have proposed a nonparametric isotropic model based on voxel distance (within the same ROI). The simulation study in Section \ref{sim_stud}, however, shows that such an assumption is overly simplistic and a nonstationary model is necessary to adequately capture the spatial structure. A wide variety of approaches have been proposed to construct nonstationary spatial processes, see \cite{gelfand:2010} for a comprehensive review. In this work, we adopt the construction of \cite{fu01}, who proposed that $\mbf{H}_{1}(v)$ is a linear combination of $L$ independent locally geometrically anisotropic processes; that is
\begin{equation}\label{laiso}
\mbf{H}_{1}(v)=\sum_{l=1}^L \mbf{H}^l_{1}(v)w_l(v),
\end{equation}
where $\mbf{H}^l_{1}(v)$ are independent, mean zero Gaussian processes across $l$ with
\begin{equation}\label{mataniso}
\text{cov}\{\mbf{H}^l_{1}(v),\mbf{H}^{l}_{1}(v')\}=\frac{1}{\Gamma(\nu_{l,r})2^{\nu_{l,r}-1}}\left[\frac{\{\mbf{R}(\mbf{v}-\mbf{v}')\}^{\top}\mbf{R}(\mbf{v}-\mbf{v}')}{\theta_{l,r}}\right]^{\nu_{l,r}}K_{\nu_{l,r}}\left[\frac{\{\mbf{R}(\mbf{v}-\mbf{v}')\}^{\top}\mbf{R}(\mbf{v}-\mbf{v}')}{\theta_{l,r}}\right],
\end{equation}
where $K_\nu$ is a Bessel function of the third kind, $\mbf{v}$ and $\mbf{v}'$ are the coordinates in 3D space of $v$ and $v'$, respectively. In other words, $\mbf{H}^l_{1}(v)$ has a Mat\'ern covariance with scale $\theta_{l,r}$, smoothness $\nu_{l,r}$ and a distance matrix $\mbf{R}$ defined as
\begin{equation}\label{anisoell}
\mbf{R}=\left(\begin{array}{ccc} \cos(\xi_{1;l,r}) & -\sin(\xi_{1;l,r}) & 0\\[7pt] \sin(\xi_{1;l,r}) & \cos(\xi_{1;l,r}) & 0 \\[7pt] 0 & 0 & 1 \end{array}\right)\left(\begin{array}{ccc} \cos(\xi_{2;l,r}) & 0 & -\sin(\xi_{2;l,r}) \\[7pt] 0 & 1 & 0  \\[7pt] \sin(\xi_{2;l,r}) & 0 & \cos(\xi_{2;l,r})  \end{array}\right)\left(\begin{array}{ccc} \ell_{1;l,r}^{-2} & 0 & 0 \\[7pt] 0 & \ell_{2;l,r}^{-2} & 0  \\[7pt] 0 & 0 & \ell_{3;l,r}^{-2}  \end{array}\right),
\end{equation}
\begin{sloppypar}
\noindent such that the isocovariance curves are ellipsoids with semi-principal axes of length $\{\ell_{1;l,r},\ell_{2;l,r},\ell_{3;l,r}\}$ and with a rotation of angle $\xi_{1;l,r}$ with respect to the $x-y$ plane and $\xi_{2;l,r}$ with respect to the $x-z$ plane. The weights $w_l(v)$ in \eqref{laiso} are the inverse distance of $v$ from the centroid of region $l$, normalized to yield unit variance. The nonstationary model in \eqref{laiso} allows for a different spatial dependence over the $L$ subregions of each ROI and thus accounts for the different local structure. The added flexibility comes at the price of an increased computational burden in the inference, as shown in Section \ref{2roi_inf}.
\end{sloppypar}

\subsection{Modeling inter-ROI dependence}

From this part, for simplicity, $\mbf{H}_2(t)$ denotes the $R \times 1$ vector of the region-specific effect instead of the $V\times 1$ vector as described in Section \ref{mod_def}.

$\mbf{H}_{2}$ is then a ROI-specific random effect with correlation matrix $\bsy{\Sigma}_2$, such that
\begin{equation}\label{cov2}
\text{cov}\{\mbf{H}_{2}(r),\mbf{H}_{2}(r')\}=\left\{
\begin{array}{llll}
(\bsy{\Sigma}_2)_{r,r'} & = & \rho(r,r') &  \text{if} \quad r\neq r',\\[7pt]
(\bsy{\Sigma}_2)_{r,r} & = & 1 & \text{if} \quad r=r',
\end{array}\right.
\end{equation}
which does not depend on the Euclidean distance between ROIs. Here, $\rho(r,r')$ is a symmetric function such that the resulting covariance is positive definite.

\section{Inference on fMRI Activation and Connectivity}\label{mod_comp}

\begin{figure}[ht]
\centering
\includegraphics[width=14cm,keepaspectratio]{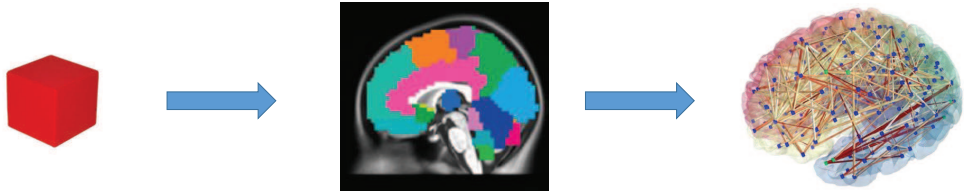}
\caption{Diagram of the three inference steps. The cube represents the voxel-specific fMRI time series. The first step considers temporal dependence of each voxel assuming spatial independence, the second the spatial dependence within each ROI assuming independence across them, and the third considers the connectivity across ROIs.}
\label{inference_scheme}
\end{figure}

Inference for the model defined in Section \ref{mod_def} needs to be performed for a data set with more than 22 million data points. Due to the extremely high dimensionality of the data, it is necessary to develop some approximations to the likelihood in order to perform inference. We propose a three-step likelihood approximation, where the second and third steps are performed conditional on the first, and assume independence of increasingly large subsets of data. In the first step, we fit a profile likelihood for each individual voxel to extract temporal dependence. The second step estimates $\mbf{H}_1(t)$ locally for each ROI, assuming each ROI is independent. The last step estimates the regional effect $\mbf{H}_2(t)$ with the entire data set. The diagram in Figure \ref{inference_scheme} shows the three inference steps. While this approach does not yield a global maximum likelihood (which is practically impossible to achieve), it has shown near-optimal results in terms of both bias and error propagation over a wide range of applications in environmental statistics \citep{ca13,ca14,ca16,ca16b,cas16}.

\subsection{Step 1: voxel-specific profile likelihood}

\begin{figure}[ht]
\centering
\includegraphics[width=14cm,keepaspectratio]{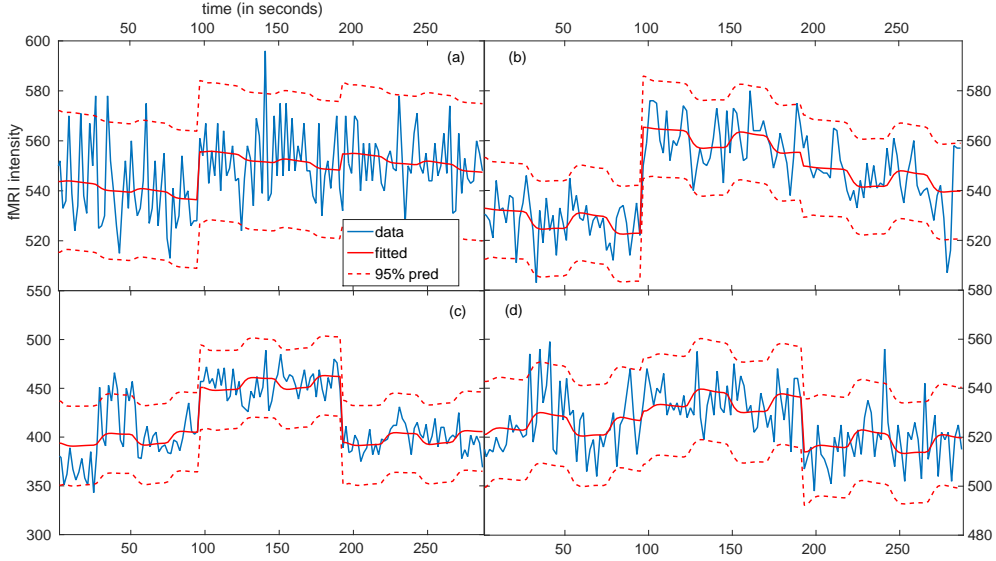}
\caption{Examples of fMRI intensity fit for four randomly selected voxels. The fitted value according to \eqref{voxel_act} (solid red) with its associated $95\%$ prediction intervals (dashed red) follow the data (solid blue). The four voxels belong to (a) middle frontal gyrus, right orbital lobe, (b) left precuneus, (c) right inferior temporal gyrus and (d) right fusiform gyrus.}
\label{voxel_fit}
\end{figure}

We initially consider models \eqref{voxel_act} and \eqref{AR2}, assuming that there is no spatial dependence. Thus, the fit of $\mbf{Y}_v=\{\mbf{Y}^{\top}_{v}(1),\ldots,\mbf{Y}^{\top}_{v}(T)\}^{\top}$ can be performed independently for each voxel via profile likelihood. If we denote the $T \times 6$ design matrix induced by \eqref{voxel_act} for every voxel as $\tilde{\mbf{X}}$, the vector of parameters for voxel $v$ as $\bsy{\theta}_v=(\phi_{1;v},\phi_{2;v},\sigma^2_v)^{\top}$, and the temporal covariance matrix induced by the AR(2) structure in \eqref{AR2} as $\mbf{K}(\bsy{\theta}_v)$, then the profile likelihood can be written as \citep{ste99}
\[
\ell(\bsy{\theta}_v;\mbf{Y}_v) = -\frac{T}{2}-\frac{1}{2}\log|\mbf{K}(\bsy{\theta}_v)| -\frac{1}{2}\mbf{Y}_v^{\top}\{\mbf{K}(\bsy{\theta}_v)^{-1}-\mbf{K}(\bsy{\theta}_v)^{-1}\tilde{\mbf{X}}\mbf{W}(\bsy{\theta}_v)^{-1}\tilde{\mbf{X}}^{\top}\mbf{K}(\bsy{\theta}_v)^{-1}\}\mbf{Y}_v,
\]
where $\mbf{W}(\bsy{\theta}_v)=\tilde{\mbf{X}}^{\top}\mbf{K}(\bsy{\theta}_v)^{-1}\tilde{\mbf{X}}$, and the mean vector $\bsy{\beta}_v$ can be obtained via generalized least squares:
\[
\hat{\bsy{\beta}}_v(\bsy{\theta}_v)=\mbf{W}(\bsy{\theta}_v)^{-1}\tilde{\mbf{X}}^{\top}\mbf{K}(\bsy{\theta}_v)^{-1}\mbf{Y}_v.
\]
Figure \ref{voxel_fit} shows the results of the fit for four randomly chosen voxels. It is apparent how the linear model \eqref{voxel_act} is able to adequately capture both the mean and the uncertainty for all voxels, including the ones that show a discontinuous change in the mean for different sessions in the experiment.

\subsection{Step 2: Estimating local effects}\label{2roi_inf}

We now consider \eqref{voxel_act} and \eqref{AR2} assuming that $\mbf{H}_1(t)$ has the spatial structure defined in subsection \ref{2roi}, while for $\mbf{H}_2$ we assume $\rho(r,r')=0$ for every $r\neq r'$ in \eqref{cov2} so that $\bsy{\Sigma}_2=\mbf{I}_R$. In other words, we assume that the process is spatially dependent within a ROI but not among ROIs.

Denote by
\[
\mbf{e}(t)= \hat{\mbf{S}}^{-1}\{\hat{\mbf{Y}}(t)-\hat{\bsy{\Phi}}_1\hat{\mbf{Y}}(t-1)-\hat{\bsy{\Phi}}_2\hat{\mbf{Y}}(t-2)\},
\]
where $(\hat{\bsy{\Phi}}_1,\hat{\bsy{\Phi}}_2,\hat{\mbf{S}})$ and $\hat{\mbf{Y}}(t)=\mbf{Y}(t)-(\mbf{I}_V \otimes \tilde{\mbf{X}})\hat{\bsy{\beta}}$ are estimated from the previous step. These residuals can then be used to estimate cov$\{\bsy{\Omega}\mbf{H}_1(t)+(\mbf{I}_V-\bsy{\Omega})\mbf{H}_2(t)\}=\bsy{\Omega}\bsy{\Sigma}_1 \bsy{\Omega}^{\top}+(\mbf{I}_V-\bsy{\Omega})(\mbf{I}_V-\bsy{\Omega})^{\top}$, and since this matrix has a block diagonal structure, the fit for each ROI can be performed independently. We thus focus on $\mbf{e}_{r}(t)$, the collection of all values of $\mbf{e}(t)$ in ROI $r$, and fit a zero mean Gaussian process with locally anisotropic covariance function \eqref{laiso} and \eqref{mataniso} to estimate $\bsy{\Sigma}_1$ (a plot of the entries of $\bsy{\Omega}$, i.e., the relative contributions of the local versus regional covariance, can be found in the supplementary material).

The choice of the number $L$ and shape of the regions in \eqref{laiso} is different for every ROI and is performed according to a model selection procedure. Each ROI is divided into $L=L_x\times L_y\times L_z$ sub-regions, where the $x$ ($y, z$) axis is divided into $L_x$ ($L_y$, $L_z$) equally spaced intervals with regions with less than 36 points merged with the largest neighboring region \citep{fu01}. For each parallelepiped $i$, the Maximum Likelihood Estimator is computed for a geometrically anisotropic Mat\'ern model assuming $\xi_{1;l,r}=\xi_{2;l,r}=0$ in \eqref{anisoell} (a model selection with further estimation of the angles was computationally infeasible), the global covariance function is estimated as a weighted sum of the local covariance functions and the Bayesian Information Criterion \citep{sc78} is computed. The optimal $L$ is chosen according to the following steps:
\begin{enumerate}
\item Start with a single region, that is with a geometrically anisotropic model with $L_x=L_y=L_z=1$.
\item Evaluate the BIC on all the neighboring configurations of $(L_x,L_y,L_z)$, that is $(L_x+1,L_y,L_z)$, $(L_x-1,L_y,L_z)$, \ldots Also, evaluate the BIC at 25 randomly drawn locations (to avoid local minima).
\item If there is a configuration with smaller BIC, redo point 2, otherwise stop.
\item Once the optimal configuration is obtained, re-estimate \eqref{anisoell} with the rotation angles.
\end{enumerate}

\begin{figure}[ht]
\centering
\includegraphics[width=14cm,keepaspectratio]{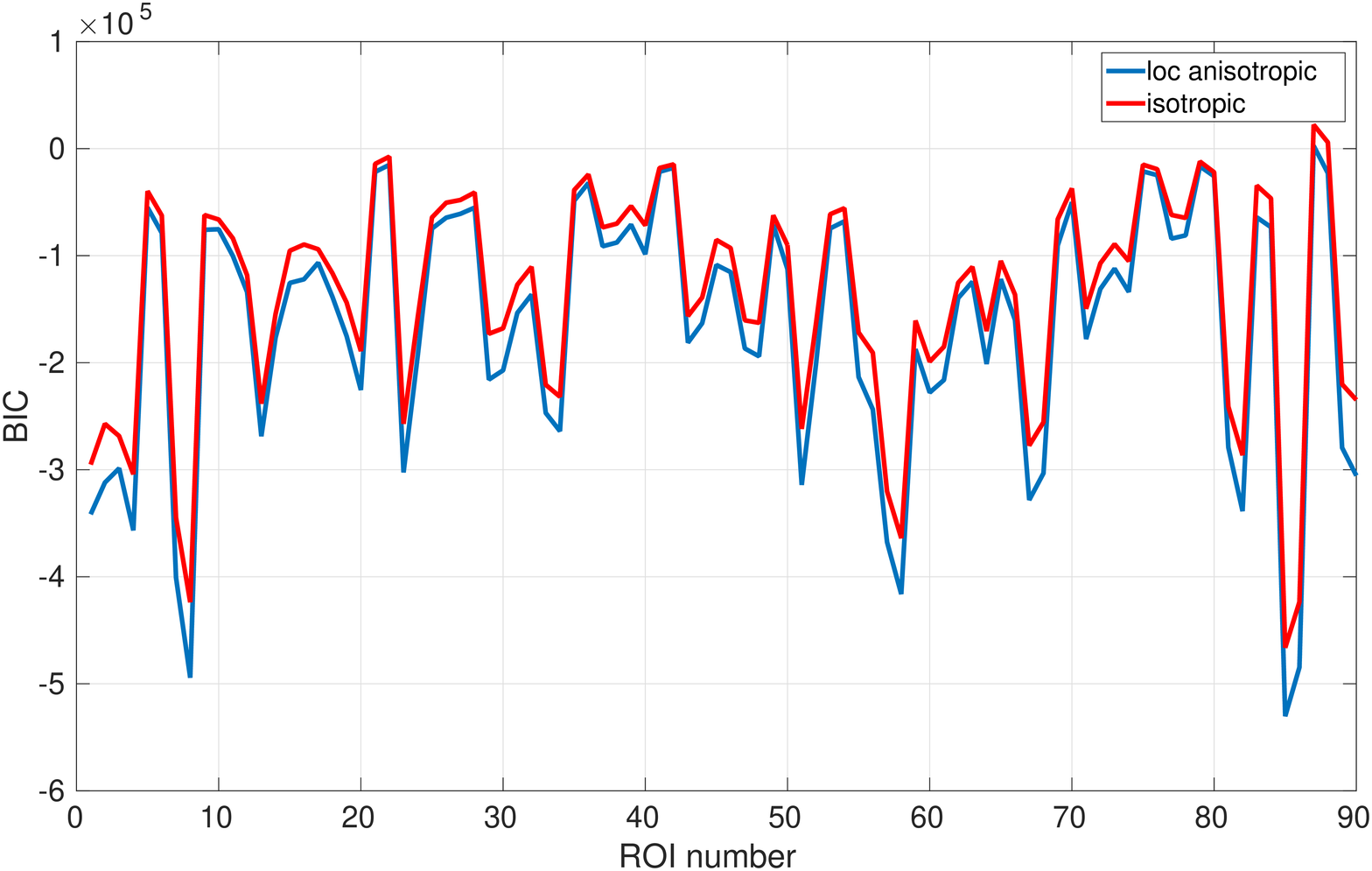}
\caption{BIC for all 90 ROIs for the isotropic model and the locally anisotropic model; the y-axis is on the $10^5$ scale.}
\label{bic_compare}
\end{figure}

A comparison of all ROIs in terms of BIC for the isotropic and locally geometrically anisotropic model is shown in Figure \ref{bic_compare}. It is apparent how the proposed model is substantially more suitable for fMRI data within the same ROI: the locally anisotropic model is on average approximately 28,000 BIC unit better than the isotropic one.

\begin{figure}[ht]
\centering
\includegraphics[width=14cm,keepaspectratio]{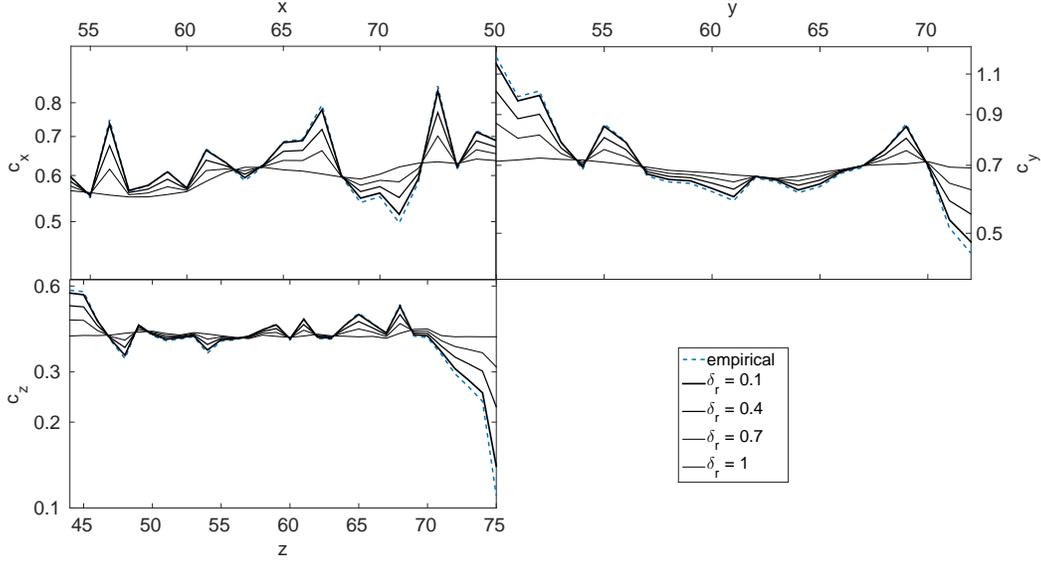}
\caption{Shrinkage for different values of $\delta_r$ in the first ROI. $c_x(\delta_r), c_y(\delta_r)$ and $c_z(\delta_r)$ are plotted against $c_x,c_y$ and $c_z$, respectively.}
\label{contrast_shrinkage}
\end{figure}

Despite its flexibility, a preliminary analysis has shown that an estimated covariance function with \eqref{laiso}, even after model selection, still leaves a considerable margin for improvement. We thus compute a new estimate of the covariance via generalized shrinkage \citep{fri08}, which allows us to estimate a new covariance matrix as a combination of the parametric model and the empirical covariance. We denote the estimated covariance resulting from the Maximum Likelihood Estimate of the $r$th block of $\hat{\bsy{\Sigma}}_1$ according to \eqref{laiso} as $\hat{\bsy{\Sigma}}^{\text{MLE}}_{1,r}$, and $\hat{\bsy{\Sigma}}_{1,r}=\frac{1}{T}\sum_{t=1}^T\mbf{e}_{r}(t)^{\top}\mbf{e}_{r}(t)$ is the sample covariance matrix of $\mbf{e}_{r}(t)$ computed from the temporal replicates. The shrunk covariance is defined as:
\[
\hat{\bsy{\Sigma}}_{1,r}(\delta_r)=(1-\delta_r)\hat{\bsy{\Sigma}}_{1,r}+\delta_r \hat{\bsy{\Sigma}}^{\text{MLE}}_{1,r},
\]
where $\delta_r\in (0,1)$ is a suitable constant, chosen so that dependence at high spatial frequencies (represented by the contrasts) matches that of $\bsy{\Sigma}_{1,r}$ within some tolerance (see Figure \ref{contrast_shrinkage}). This is obtained with the following steps:
\begin{itemize}
\item The contrast
\[
\begin{array}{lll}
c_x & = & \frac{1}{|N_x|}\sum_{(y,z)\in S_x} \{(\hat{\bsy{\Sigma}}_{1,r})_{(x,y,z),(x,y,z)}+ (\hat{\bsy{\Sigma}}_{1,r})_{(x+1,y,z),(x+1,y,z)}-2(\hat{\bsy{\Sigma}}_{1,r})_{(x+1,y,z),(x,y,z)}\}
\end{array}
\]
is computed, where $N_x$ is the set of all pairs $(y,z)$ such that the points $(x,y,z)$ and $(x+1,y,z)$ belong to the grid. Similarly we define $c_y$ and $c_z$.
\item A smoothing spline with penalization parameter $p$ is fit to $c_x, c_y$ and $c_z$, which we call $c^{p}_x, c^{p}_y$ and $c^{p}_z$, respectively.
\item The parameter $\delta_r$ generates $\bsy{\Sigma}_{1,r}(\delta_r)$ such that
\[
\begin{array}{lll}
c_x(\delta_r) & = & \frac{1}{|S_x|}\sum_{(y,z)\in S_x}\{(\hat{\bsy{\Sigma}}_{1,r}(\delta_r))_{(x,y,z),(x,y,z)}+(\hat{\bsy{\Sigma}}_{1,r}(\delta_r))_{(x+1,y,z),(x+1,y,z)}\\[7pt]
              &  & -2(\hat{\bsy{\Sigma}}_{1,r}(\delta_r))_{(x+1,y,z),(x,y,z)} \}
\end{array}								
\]
Similarly we define $c_y(\delta_r)$ and $c_z(\delta_r)$. We choose $\delta_r$, such that $\|c^{p}_x-c_x(\delta_r)\|^2_2+\|c^{p}_y-c_y(\delta_r)\|^2_2+\|c^{p}_z-c_z(\delta_r)\|^2_2$ is minimized.
\end{itemize}

Thus, it is possible to control the spatial structure via the penalization parameter $p$. Figure \ref{contrast_shrinkage} shows an example of the shrinkage to the empirical covariance for a fixed ROI and different values of $\delta_r$ in terms of their spatial contrasts. For our analysis we choose a penalty term of $p=0.3$, because it allows for some flexibility in the pattern structure, and because values of $p$ in this neighborhood have yielded qualitatively indistinguishable results.

\subsection{Step 3: estimating the regional effects}

\begin{figure}[ht]
\centering
\includegraphics[width=14cm,keepaspectratio]{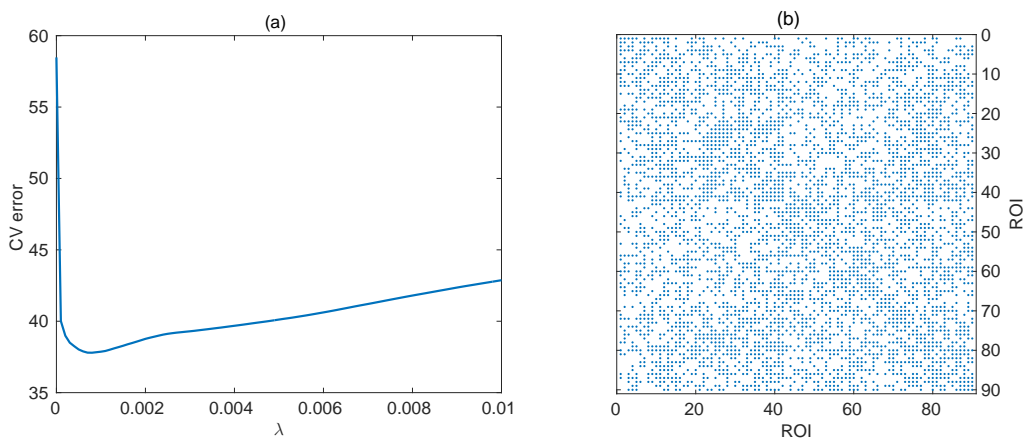}
\caption{Graphical LASSO for brain connectivity. (a) Sum of squares cross-validation error versus different choices of penalty in \eqref{glasso}, (b) plot of all nonzero elements in $\mbf{W}_{\text{brain}}$ for the optimal $\hat{\lambda}$.}
\label{connectivity_plot}
\end{figure}

Conditional on the estimated covariance structure in the previous steps, we estimate the covariance structure of $\mbf{H}_2(t)$, that is the entries of $\bsy{\Sigma}_2$. If we denote as $\bar{\mbf{e}}(t)=\{\bar{\mbf{e}}^{\top}_{1}(t),\ldots,\bar{\mbf{e}}^{\top}_{R}(t)\}^{\top}$, and $\bar{\mbf{e}}_{r}(t)$ the average of $\mbf{e}(t)$ for ROI $r$, $\mbf{A}=\frac{1}{T}\sum_{t=1}^T \bar{\mbf{e}}(t)^{\top}\bar{\mbf{e}}(t)$ provides a nonparametric estimation of $\bsy{\Sigma}_2$, the correlation matrix for the ROI specific effects. However, this would require estimating $R(R-1)/2$ entries, with no insight on the connectivity patterns induced by the experiment, that is, which ROIs are significantly connected. We choose to estimate the $R \times R$ inverse covariance $\mbf{W}_{\text{brain}}$ of $\mbf{H}_2(t)$, and to impose an $\ell^1$ constraint on the number of nonzero entries of the inverse correlation matrix. This penalized likelihood approach on the inverse covariance was proposed by \cite{yua07} as an application of the maxdet problem \citep{van98}; a faster approach for high dimensional covariance estimation, the graphical LASSO \citep{fri08}, has become widely popular in recent years, also in the context of neurological data \citep{var10,cri12}. The main idea is to rewrite the loglikelihood for $\bar{\mbf{e}}(t)$
\[
-\frac{T}{2}\log(2\pi)+\frac{1}{2}\log|\mbf{W}_{\text{brain}}|-\frac{1}{T}\sum_{t=1}^T \bar{\mbf{e}}(t)^{\top}\mbf{W}_{\text{brain}}\bar{\mbf{e}}(t)
\]
as a disciplined convex problem, and to further impose an $\ell^1$ penalty on the number of nonzero entries of $\mbf{W}_{\text{brain}}$:
\begin{equation}\label{glasso}
-\frac{T}{2}\log(2\pi)+\frac{1}{2}\log|\mbf{W}_{\text{brain}}|+\text{tr}(\mbf{W}_{\text{brain}}\mbf{A})+\lambda \sum_{r\neq r'}|(\mbf{W}_{\text{brain}})_{r,r'}|.
\end{equation}
This allows us to obtain an estimate of the inverse correlation that is sparse, and consequently interpretable results for ROI connectivity. The optimal $\hat{\lambda}$ is chosen by cross-validation \citep{fri08}: the inverse correlation is evaluated with \eqref{glasso} for $90\%$ of the data. Then, each ROI is predicted on the remaining $10\%$ of the data by leave-one-out cross-validation. Figure \ref{connectivity_plot}a shows a plot of the error sum of squares against $\lambda$. A small penalty results in severe overparametrization and as the penalty increases, the estimated $\mbf{W}_{\text{brain}}$ is closer to the diagonal matrix and results in a suboptimal fit. In the supplement, the file movie\_glasso.avi shows how the sparsity increases and where it occurs as $\lambda$ increases. Figure \ref{connectivity_plot}b shows all the nonzero elements ($53\%$ of the total number of entries) of $\mbf{W}_{\text{brain}}$ for the optimal $\hat{\lambda}=8\times 10^{-4}$. 

Once the optimal sparse inverse correlation is computed, it is possible to infer the structure of brain connectivity from the structure of  the graph, as will be discussed in Section~\ref{s5disc}.

\subsection{Computational considerations}

Besides providing a flexible framework to allow inference of a complex model with more than 22 million data points, this multi-resolution scheme also allows for an extensive use of distributed computing to achieve a scalable inference. In step 1 the code is parallelized so that inference is performed independently for every voxel time series. In step 2, model \eqref{laiso} with \eqref{mataniso} (and similarly the simpler models in the simulation studies) can be estimated independently for each core in a cluster, allowing as many as $R$ independent simultaneous estimations. It is also possible to focus on a single ROI and parallelize the model selection algorithm detailed in subsection \ref{2roi_inf}, but this approach yielded a suboptimal performance, as the communication overhead across processors generated significant latency especially with small-sized ROIs.

Steps 1 and 3 were performed on a workstation with two twelve-core Intel Xeon E5-2697 v2 (at nominal frequency 2.7Ghz) and 200 Gb of RAM, which required approximately 4 hours for step 1 (step 3 can be performed within seconds). Step 2 was performed on a fully dedicated cluster with 27 nodes (each with 20 cores and 128 Gb of RAM) and required approximately 3 days. All the likelihood inferences were performed in MATLAB with the Nelder--Mead minimization algorithm.

\subsection{Testing for voxel-specific activation}\label{testvo}

\begin{figure}[ht]
\centering
\includegraphics[width=8cm,keepaspectratio]{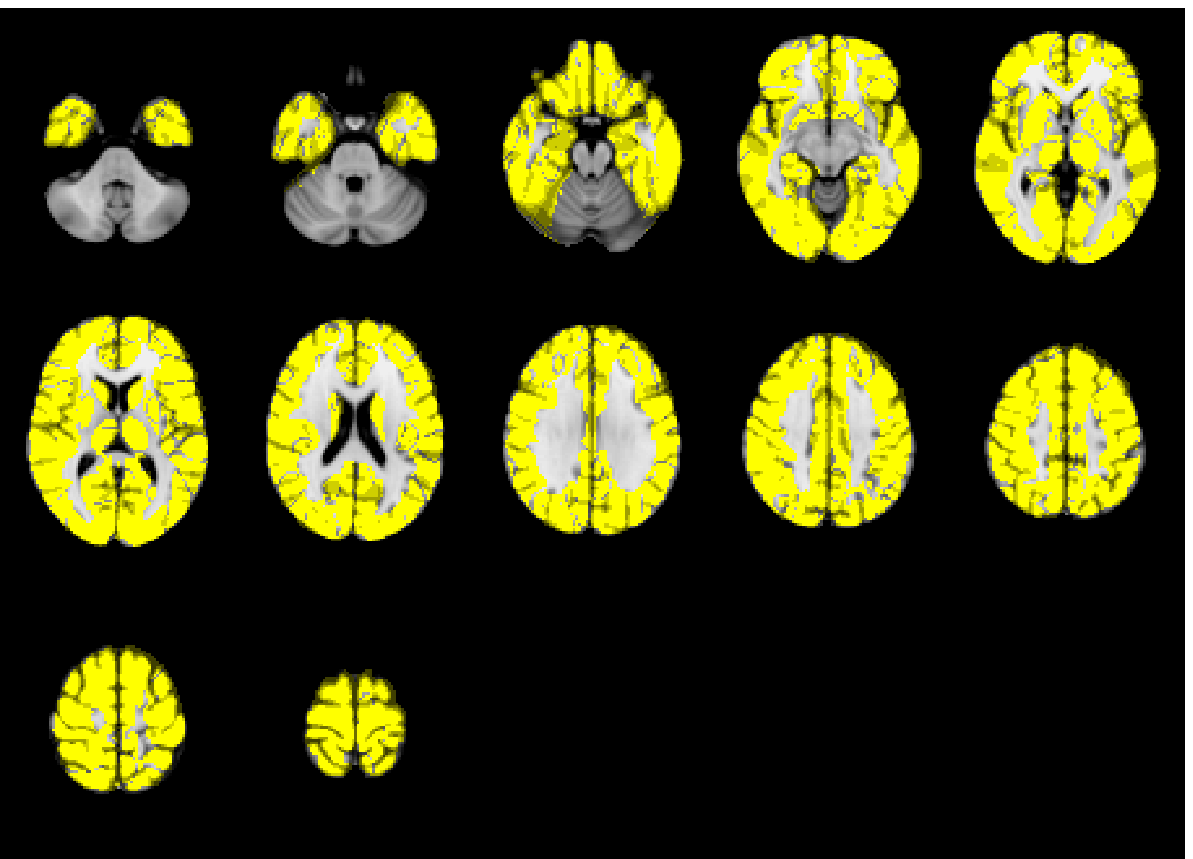}
\vfill
\includegraphics[width=8cm,keepaspectratio]{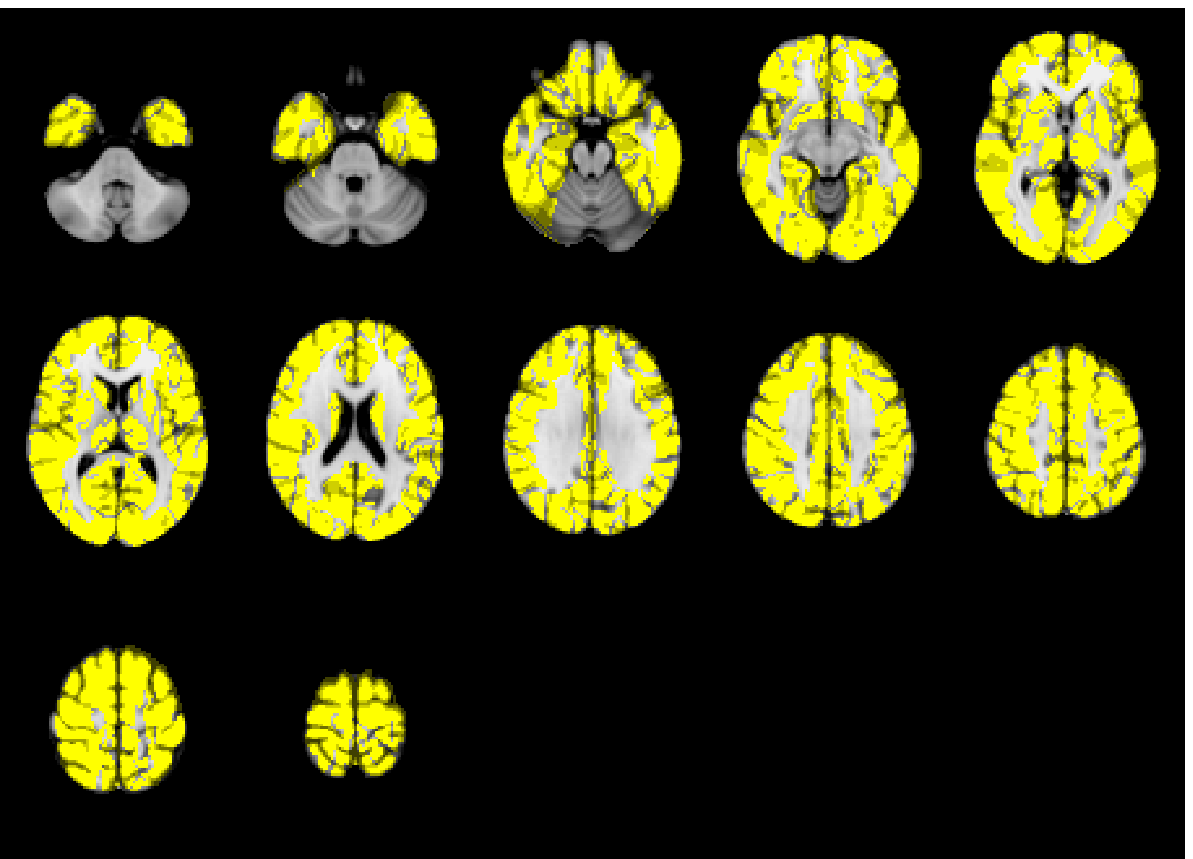}
\caption{Activated voxels at a False Discovery Rate of $5\%$ for 10 equally spaced slices on the $z$-axis (top) assuming independent voxels and (bottom) with the proposed spatial dependence.}
\label{plot_activated_roi}
\end{figure}

Activation for each voxel can be tested by combining a fine-scale estimation of the dependence structure with the definition of a voxel-specific mean structure. We assume that $\bsy{\beta}_{0}, \ldots, \bsy{\beta}_{J+1}$ are fixed and equal to the estimated values obtained from the fit of the temporal and mean parameter in \eqref{voxel_act} assuming spatial independence, while we re-estimate $\bsy{\beta}_{J+2}$ and $\bsy{\beta}_{J+3}$, for which we want to test H$_0$: $\bsy{\beta}_{J+2}-\bsy{\beta}_{J+3}=0$ versus H$_1:\bsy{\beta}_{J+2}-\bsy{\beta}_{J+3}\neq 0$. We allow $\bsy{\beta}_{J+2}$ to be smoothly varying in space, thus allowing to borrow strength across neighboring locations and yielding a more accurate test for activation. We assume that the spatial variation is deterministic and parametrized by Fourier coefficients: for each ROI $r$, we have 
\begin{equation}\label{spc}
\bsy{\beta}^r_{J+2}(x,y,z)=\sum_{n=1}^N \sum_{ b \in \{x,y,z\}} a^r_{b,n,1} \cos(2\pi b n/d_b^r)+a^r_{b,n,2} \sin(2\pi b n/d_b^r),
\end{equation}
where $x,y$ and $z$ are the (normalized) coordinates over the three axes, $d^r_x,d^r_y$ and $d^r_z$ are the maximum lengths across the three axes. To avoid identification issues, we set $\bsy{\beta}_{J+3}$ constant across the ROI. We thus need to estimate $\{a^r_{b,n,1},a^r_{b,n,2}\}$ for each harmonic $n=1,\ldots, N$, for each coordinate $b \in \{x,y,z\}$ and for each ROI $r=1,\ldots, R$. As in the previous sections, each ROI can be estimated independently and the code can be easily parallelized, but the computational and memory demand for each likelihood evaluation allowed only to fit the model for $N=1$. Figure \ref{plot_activated_roi} shows the activated voxels for each ROI with False Discovery Rate \citep{ben95} at $5\%$ assuming independence and spatial dependence. The results for particular areas will be discussed in Section \ref{s5disc}.

Alternatively, the spatial variation could be also modeled hierarchically, i.e. by assuming $\bsy{\beta}_{J+2}\sim \mathcal{N}(\mbf{m},(\lambda\mbf{Q})^{-1})$ as a latent process with mean $\mbf{m}$, precision matrix $\lambda\mbf{Q}$ with (fixed or random) parameter $\lambda$. However, the computational burden of evaluating the likelihood for the entire ROI would make the inference considerably more challenging.

\section{Simulation Studies}\label{sim_stud}

To further support our choice of a locally anisotropic model versus the simpler existing alternatives described in subsection \ref{2roi_inf}, we perform two simulation studies: one focused on an activation test and one focused on spatial interpolation.

In the first study, we focus on the improved performance of a locally anisotropic model (\textit{l-aniso}) against a general linear model (denoted \textit{glm}), an isotropic model (\textit{iso}) and an anisotropic model (\textit{aniso}) for a single ROI. For each ROI, we perform $100$ simulations, where the true spatial covariance is the sample covariance obtained from the fMRI data of the subject in the case study after detrending with Ordinary Least Squares in time. This approach ensures that the performance of each method is compared against a true model that is as close as possible to the original data set. For computational reasons, whenever an ROI has more than 1000 voxels, a random sample of this size (the same for every simulation) is considered for the analysis. We assume a common mean across the ROI with no time or session effect and only the hemodynamic response terms in the mean structure, which we denote as $\beta_1$ and $\beta_2$. The common mean does not allow a straightforward comparison with the results in subsection \ref{testvo}, but it does significantly reduce the considerable computational burden and allows for a closer comparison with the similar simulation studies in \cite{ka12}. For each simulation, $\hat{\beta}_1-\hat{\beta_2}$ and $\hat{\text{var}}(\hat{\beta_1}-\hat{\beta_2})$ are estimated according to the four models and a test is performed to determine the presence of activation, i.e. if $\beta_1\neq\beta_2$. Model selection for the locally anisotropic model is avoided because it would require several days per simulation on a large computational facility; the geometry of the anisotropic subsets of the brain is obtained from the model selection step in subection \ref{2roi_inf}.

In the first study, we assume that the ROI is not active, i.e. $\beta_1=\beta_2$ and we compare the false positives at a $5\%$ confidence level out of the 100 simulations. Table \ref{sim1res} shows the results, where the first three rows represent the number of false positives for three randomly drawn ROIs and the last row computes an average across all ROIs. The assumption of no spatial dependence in \textit{glm} implies a very high number of false positives, as was previously reported by \cite{ka12}. Even with the simple isotropic model \textit{iso}, the assumption of spatial dependence brings dramatic improvements in the accuracy of the test. The models \textit{aniso} and \textit{l-aniso} bring a further improvement, although it comes at the expense of a longer computational time. It is also remarkable how these results are markedly larger than the nominal $5\%$ rate of false positive, thus indicating how the nonstationarity dependence within ROI is very complex and more sophisticated models could further improve the results. A fully nonparametric approach of estimating the empirical covariance (results not shown) proved considerably worse, yielding nearly 100$\%$ false positives, indicating that some parametric description of the nonstationarity is needed. In the supplementary material, we show the power curves for this study.

\renewcommand\arraystretch{1.3}
\begin{table}[tbhp]
\caption{{\small Percentage of false positives assuming no activation for three randomly chosen ROIs (first three rows, see Table S1 for the abbreviation) and the mean across ROIs for the four models (last row). }}\label{sim1res}
\centering
{\footnotesize \begin{tabular}{|c|c|c|c|c|}
\hline
ROI   & \textsl{general linear model} & \textit{isotropic} & \textit{anisotropic} & \textit{locally anisotropic} \\ \cline{1-5}
 Superior occipital gyrus, Right  & 80 & 26 & 28 & 13 \\ \cline{1-5}
 Parahippocampal gyrus, Right  & 72 & 14 & 12 & 11 \\ \cline{1-5}
 Orbital Superior frontal gyrus, Left  & 64 & 35 & 31 & 27 \\ \cline{1-5}
mean   &  78.7 &  31.9 & 28.3 & 26.3  \\ \cline{1-5}
\end{tabular}}
\end{table}

In the second study, we compare the effect of the four models on interpolation. In the setting of the previous studies, for each simulation we remove 50 random voxels (the same across all simulations for the same ROI), we interpolate their values with kriging and compute the Root Mean Squared Error (RMSE) with the true value. The results in Table \ref{sim3res} show how the independence assumption is largely inadequate, and how \textit{iso}, \textit{aniso} and \textit{l-aniso} perform incrementally better as they yield interpolated values closer to the true simulated data. As in the first simulation study, a smaller RMSE for \textit{aniso} and \textit{l-aniso} comes at the cost of a much more computationally challenging estimation. Although the difference in performance between \textit{aniso} and \textit{l-aniso} seems small, note that the model selection was not performed for every simulation because it would have been too computationally demanding. Thus, further improvement could be expected if the regions of anisotropy were not predefined from the real data set.

\renewcommand\arraystretch{1.3}
\begin{table}[tbhp]
\caption{{\small RMSE$\times 10^3$ for 50 randomly removed points for three randomly chosen ROIs (first three rows, see Table S1 for the abbreviation) and the mean across ROIs for the four models (last row). }}\label{sim3res}
\centering
{\footnotesize \begin{tabular}{|c|c|c|c|c|}
\hline
ROI   & \textsl{general linear model} & \textit{isotropic} & \textit{anisotropic} & \textit{locally anisotropic} \\ \cline{1-5}
 Superior occipital gyrus, Right  & 6.97 & 1.09 & 1.05 & 1.04 \\ \cline{1-5}
 Parahippocampal gyrus, Right  & 6.89 & 1.05 & 0.94 & 0.91 \\ \cline{1-5}
 Orbital Superior frontal gyrus, Left  & 6.93 & 0.99 & 0.99 & 0.95 \\ \cline{1-5}
mean   &  6.66 & 1.10 & 1.01 & 0.99  \\ \cline{1-5}
\end{tabular}}
\end{table}

\section{Discussion of the Results}\label{s5disc}

The model selection procedure suggests that a locally anisotropic model outperforms the isotropic model uniformly across ROIs (see Figure~\ref{bic_compare} and Tables~\ref{sim1res} and \ref{sim3res}). Due to the computational complexity, previous models (e.g., \cite{ka12,ka13}) have simply assumed an isotropic structure within each ROI. However, our method suggests non-stationary behavior, even within an ROI, which means complexity of the spatial covariance structure that requires more sophisticated modeling.
In particular, the degree of correlation between voxels may vary
as a function of their Euclidean distance, although this correlation may differ depending
on the exact location of these voxels. This difference could have a significant impact
on inference (e.g., testing for activation) and hence must be properly specified when
computing the test statistic.

One of the aims of our analysis is to determine the voxels that are activated during the hand-grasping task. In Figure~\ref{plot_activated_roi}(a-b), we show the contrasts for the analysis that ignores spatial correlation (in black) and the analysis that incorporates local anisotropy (in blue) for the left supplementary motor area (SMA-L), while in the supplementary material we report (Table S2) the percentage of all activated voxels according to the two models for all ROIs. 

We note that the activation patterns differ in many respects. Firstly, the contrasts estimation for the spatio-temporal model is considerably more conservative (as shown in Figure~\ref{plot_activated_roi}(a-b)), and on average there is a smaller number of voxels with significantly different BOLD activation between the hand-grasping and rest conditions ($92.8\%$ for the general linear model and $84.8\%$ for the spatio-temporal model, from Table S2). It is very likely that here, the independence model flagged as significant a number of voxels that in fact do not display a differential BOLD response level. Secondly, among all ROIs, SMAL-L had the highest percentage of voxels with significantly greater amplitude of the BOLD response during the hand-grasping task compared to the resting state (see Table S2) under the spatio-temporal model. This result was expected, since the subject was performing a motor task using his right (dominant) hand. Besides, the corresponding ROI in the right hemisphere, the right supplementary motor area (SMA-R) has the third largest proportion of activated voxels at $96\%$. Under the independence model, SMA-L and SMA-R ranked 10th and 6th respectively.

\begin{figure}[ht]
\centering
\includegraphics[width=17cm,keepaspectratio]{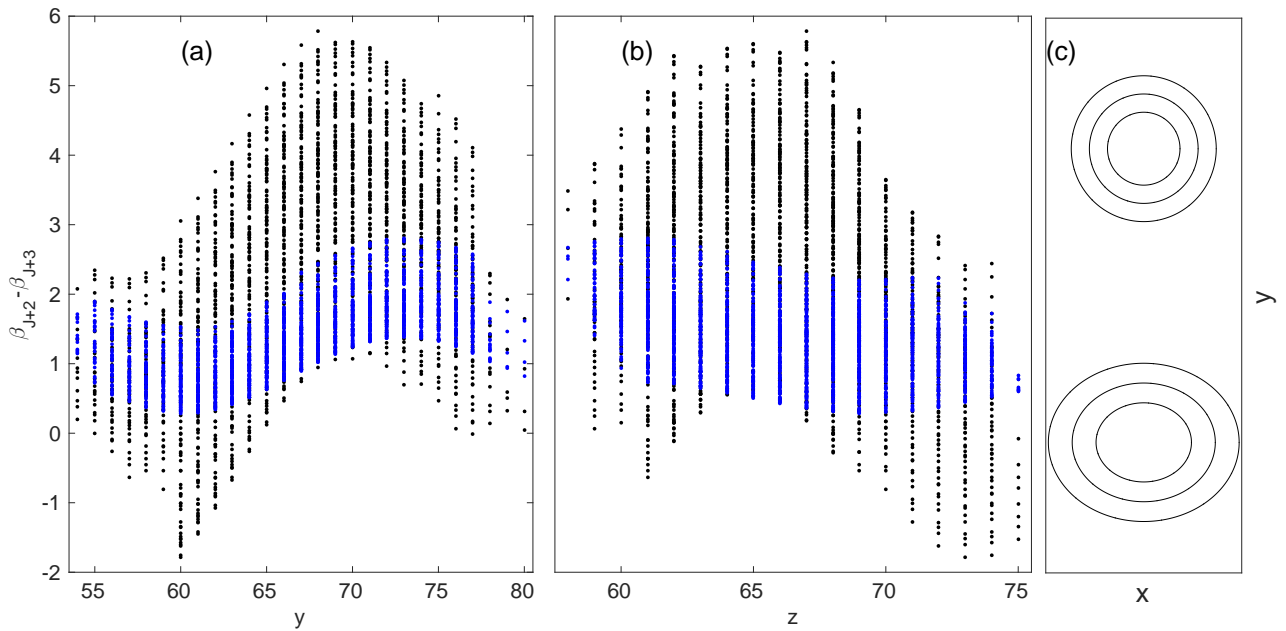}
\caption{Activation contrasts against the $y$ (a) and $z$ axis (b) for SMA-L. Black dots represent the contrast according to the model with spatial independence, while the blue dots represent the spatio-temporal model. (c) The contour of the estimated covariance function for two points in the anterior (top) and posterior (bottom) part of the ROI.}
\label{plot_activation_laniso_sma}
\end{figure}

We also identified an unexpected result using the independence model that carries no neurophysiological justification: right precentral area was flagged as the most significant region. This result was quite unexpected since the right area is indicated as more active, as it should be in a left-handed subject, and the SMA-L is only ranked 10th in terms of active voxels. Also, the cuneus left (CUN-L) had the second largest percentage of voxels with higher BOLD amplitude during the motor task, and this region is not located in the primary motor cortex. This underscores the fact that ignoring spatial covariance in the analysis could produce unexpected results that are likely incorrect since the task is purely motor and would not require higher cognitive processing.

In addition to examining activation, the proposed procedure can also be used to study 
complex localized spatial covariance. As noted, the BIC model selection metric chose the locally anisotropic 
model over the isotropic model. This concept of local anisotropy is not just a theoretical 
construct. Firstly, we need the spatial covariance to be properly specified to give us confidence about the level and power of our testing procedure. Next, the model provides us with information on the strength of correlation between pairs of voxels and how it may vary,
depending on whether the pair is located on the anterior or posterior neighborhood of
the ROI. For example, in Figure \ref{plot_activation_laniso_sma}c the anterior part of 
SMA-L, has a the correlation structure that is more circular than ellipsoidal (meaning there is no preferred direction in the dependence structure), while the opposite is true of the anterior right part of SMA-L.  
This finding is very interesting for our stroke collaborators who will look into confirming these findings using other types of data modalities and modeling techniques (e.g., probabilistic tractography 
and fractional anistropy/mean diffusivity in diffusion tensor imaging). However, this is beyond the current scope of this paper. 

The task of modeling and understanding connectivity is intertwined with activation and of particular interest in stroke patients because neurons from other regions activate at a higher than normal level to compensate for reduced activation in regions with injury. The estimated inverse covariance matrix revealed a number of interesting connections between a few pairs of ROIs. Using a high threshold at $\lambda=0.01$, the pair of regions that survived the stringent threshold, indicating the strongest pairwise direct dependence, is left frontal operculum and left rolandic operculum. 
The left frontal operculum refers to the small region in the frontal lobe that overlies the rostrodorsal portion of the insula in primates. 
\cite{ale90} found 9 cases of aphasia (broadly defined as a 
difficulty with speech and writing) following lesions in the region of the left frontal 
operculum. Moreover, in \cite{ton81}, two cases of articulatory difficulty were associated with 
lesions in the rolandic operculum. It is interesting that these two ROIs are almost 
adjacent, which partly explains how damages to these regions result in similar 
symptoms of aphasia. This suggests that the strongest direct link between ROIs 
may not be due to these regions having shared functional dependence because these 
ROIs are not shown to be implicated in motor task activity, however, this can be explained 
by the actual anatomical proximity.

\section{Conclusions}\label{s6conc}

This work addressed the issue of enhancing the detection of activation and connectivity in fMRI data by explicitly modeling spatial dependence.  Motivated by the need to develop flexible models for enhancing detection of neurological patterns in the recovery of a patient suffering stroke, this work provides a general framework for whole-brain modeling of a single patient. Ultimately, this work will be extended to multiple patients to allow for comparisons across subjects in a follow-up investigation. 

The multi-resolution model introduced allows for spatially varying coefficients and local nonstationarity within ROIs in the error structure. We have demonstrated with numerical studies how isotropy within an ROI, a widespread simplifying assumption, is largely inappropriate for fMRI data even for small ROIs, as it leads to suboptimal tests for activation. Compared to the current methodology, our proposed model showed clear improvements. 
In particular, numerical studies suggest the nonstationarity is complex and a locally anisotropic model is not fully adequate. Future work will investigate the use of other constructions for nonstationary processes such as the one proposed in \cite{pac06} or multiresolution models with random coefficients with sparse dependence \citep{nyc15}. It is expected that more flexible models for local nonstationarity will result in a type I error closer to the nominal value.  

This model shares common features from the two-stage Bayesian hierarchical approach introduced in \cite{Bowman2007} and the spatio-spectral mixed model in \cite{ka12} in that it aims at modeling spatial dependence directly instead of mitigating its effect via spatial smoothing. However, it provides a finer spatial scale information on activation, which is attained at the price of a substantially increased computational burden, requiring a tailored inference scheme that fully exploits parallelization. In addition to the advantage of describing finer scale information, voxel-specific (rather than an ROI-specific) activation bypasses the problem of determining how many active voxels comprise an active ROI \citep{ka12}. The connectivity is captured via the ROI-specific random effect as in \cite{ka12}, but the model we propose is more appealing, as graphical LASSO yields undirected graphs that are interpretable and partially avoids the overparametrization of the empirical covariance.



\baselineskip=10pt

\bibliographystyle{asa}
\bibliography{citations}

\end{document}